\documentstyle[aps,twocolumn,prl,epsf,floats]{revtex}

\begin{document}
\twocolumn[\hsize\textwidth\columnwidth\hsize\csname @twocolumnfalse\endcsname

\title{Orbital Magnetism in Ensembles of Parabolic Potentials} 
\author{M.O. Terra, M.L. Tiago and M.A.M. de Aguiar} 
\address{ Instituto de F\'{\i}sica `Gleb Wataghin',             
Universidade Estadual de Campinas, Unicamp,\\ 
13081-970, Campinas, S\~ao Paulo, Brasil\\
}

\maketitle

\begin{abstract}
 We study the magnetic susceptibility of an ensemble of non-interacting
electrons confined by parabolic potentials and subjected to a
perpendicular magnetic field at finite temperatures. We show that the 
behavior of the  average susceptibility is qualitatively different
from that of billiards. When averaged over the Fermi energy 
the susceptibility exhibits a large paramagnetic response only
at certain special field values, corresponding to comensurate classical
frequencies, being negligible elsewhere. We derive approximate
analytical formulae for the susceptibility and compare the results with 
numerical calculations.
\end{abstract}


% 05.30.Ch Quantum ensemble theory
% 03.65.Sq Semiclassical theories and applications
% 73.20.Dx Electron states in low-dimensional structures (superlattices, 
%                  quantum well structures and multilayers)
% 73.23.-b Mesoscopic systems
]
\newpage
%%%%%%%%%%%%%%%%%%%%%%%%%%%%%%%%%%%%%%%%%%%%%%%%%%%%%%%%%%%%%%%%%%%%%%%%%
%%%%%%%%%%%%%%%%%%%%%%%%%%%%%%%%%%%%%%%%%%%%%%%%%%%%%%%%%%%%%%%%%%%%%%%%%

The interest in the magnetic properties of ensembles of mesoscopic
systems has increased considerably in the last years 
\cite{ullmo,report,bara} . The main motivation for the theoretical
investigations recently carried out is the experimental results
obtained by Levy et al \cite{levy} for an ensemble of square billiards
in the ballistic regime. It is now understood that, when averaged over a 
large ensemble of similar systems, the magnetic susceptibility $\chi$ of 
regular systems is enhanced with respect to the Landau susceptibility 
$\chi_L$ due to the coherent contribution of families of periodic orbits. 
For chaotic systems $\chi$ is usually small, of the order of $\chi_L$, but
bifurcations might play an important role in increasing the susceptibility
\cite{prado1}. Also, for square billiards, the averaged susceptibility is 
always paramagnetic at low magnetic fields. This behavior seems to be 
generic of regular billiards \cite{ullmo,report}.

The purpose of this letter is to study the magnetic susceptibility 
of an ensemble of noninteracting two-dimensional electron gas confined by  
parabolic quantum wells at finite temperatures. We show that the behavior 
of $\chi$ as a function of the magnetic field $B$ and Fermi energy $\mu$ is 
very different from that of billiards when ensemble averages are considered. 
The main reason for the this strong difference resides on the resonances 
exhibited by the system as the magnetic field is varied. The susceptibility 
of a single harmonic oscillator in a magnetic field has been considered 
before by Prado et al \cite{prado} and Nemeth \cite{nemeth}. In this work
we derive simple analytic expressions for $\chi$ which
are directly amenable of ensemble averages.

In the canonical ensemble the magnetic susceptibility per particle
$\chi  = -(1/N) \; \partial^2 F/\partial B^2$ measures the 
sensitivity of the Helmholtz free energy $F$ to the magnetic field $B$.
For a non-interacting model $F$ can be computed exactly if the
number of particles $N$ and the temperature $T$ are not too large 
\cite{calcf}. However, for
large $N$ and temperatures of the order of the the mean 
level spacing (in units of Boltzmann constant $k_B$), $\chi$ can be computed 
quite accurately by using an auxiliary grand-canonical potential where
the average number of particles is kept fixed by properly adjusting the
chemical potential for each value of the magnetic field  \cite{altshuler}.
The grand-canonical potential is given by

\begin{equation}
V = -\frac{1}{\beta} \int dE \; \rho (E) \; \ln(1+e^{\beta(\mu -E)})
\label{gran}
\end{equation}

\noindent where $\rho$ is the density of states, $\beta=1/k_{B}T$ and 
$\mu$ is the chemical potential. In the semiclassical
limit $\rho$ can be separated into a mean term $\rho^0$ plus 
oscillatory contributions $\rho^{osc}$ that are usually written in terms of
period orbits \cite{gutz}. After substituting $\rho=\rho^0+\rho^{osc}$ in
Eq.(\ref{gran}), the grand-canonical potential also separates
into $V^0(\mu)+V^{osc}(\mu)$.  Following Ullmo et 
al \cite{ullmo} we define a mean chemical potential $\mu^0$ from 
$N=\int {\rm dE} \; \rho(E) f(E-\mu) = \int {\rm dE} \; \rho^0(E) f(E-\mu^0)$ 
where $f(x)=1/(1+e^{\beta x})$ is the Fermi-Dirac distribution function.
Using the thermodynamical relation $F(N)=V(\mu) + \mu N$ with
$\mu=\mu(N)$ obtained from the equations above, it can be shown 
\cite{altshuler} that, in the semiclassical limit, the Free Energy can be 
written as a sum of three terms, $F = F^0 + \Delta F^1 + \Delta F^2$, where
$F^0 = V^0(\mu^0) + \mu^0 N$ does not depend on the magnetic
field $B$, $\Delta F^1 = V^{osc}(\mu^0)$ and
\begin{equation}
\Delta F^2 = \frac{1}{2 \rho^0(\mu^0)} \left[\int {\rm dE}\; \rho^{osc}(E)
f(E-\mu^0) \right]^2 \; \; .
\end{equation}
In the case of a parabolic confinement, the Hamiltonian of an electron
of charge $e$ and effective mass $m^*$ can be written directly in
terms of action variables as  \cite{schuh} 
$H({\bf I})={\bf \Omega}\cdot {\bf I}$ with 
$\Omega_1 = \frac{1}{2}(\xi_+ + \xi_-)$, 
$\Omega_2 = \frac{1}{2}(\xi_+ - \xi_-)$ and
$\xi_{\pm} = \sqrt{(\omega_1 \pm \omega_2)^2 + e^2B^2/{m^*}^2}$.
The quantum mechanical energy levels are therefore given by $E_{k_1 k_2} = 
\hbar \Omega_1(k_1 + 1/2) + \hbar \Omega_2(k_2+1/2)$. To compute the
$\rho^0$ and $\rho^{osc}$  we follow Berry and
Tabor \cite{berrytabor} and write the semiclassical density of states as
\begin{equation}
\rho(E) = 2 \sum_{\bf m} \delta(E-H({\bf I}={\bf m}+{\bf 1}/2)) \; .
\end{equation}

\noindent where ${\bf m}=(m_1,m_2)$ and the factor $2$ takes care of
spin degeneracy. Using the Poisson sum formula we find 
$\rho^0(E)=2 E/(\hbar^2 \omega_1 \omega_2)$ and
\begin{equation}
\rho^{osc}=\frac{2}{\hbar^2} {\sum_{\bf m}}^{'} e^{-i \pi (m_1+m_2)} 
\rho_{\bf m}
\end{equation}
with
\begin{equation}
\rho_{\bf m}=\int {\rm d} \xi \frac{e^{2 \pi i {\bf m}
\cdot {\bf I}(\xi)/\hbar}}{|{\bf \omega}({\bf I}(\xi))|}
\label{rom}
\end{equation}
where $\xi$ varies along the energy surface in the $(I_1,I_2)$ plane of
the action variables. The prime above the summation symbol means that
the term $(m_1,m_2)=(0,0)$ (responsible for $\rho^0$) is excluded. 
Explicitly we get 
$I_1=E/\Omega_1-\xi \Omega_2/\Omega$ and $I_2=\Omega_1 \xi/\Omega$ for 
$0 \leq \xi \leq E \Omega/(\Omega_1 \Omega_2)$ with 
$\Omega \equiv \sqrt{\Omega_1^2+\Omega_2^2}$. For generic Hamiltonians 
the integration over $\xi$ can performed within the stationary
phase approximation and gives a semiclassical expression for 
$\rho^{osc}$. In the present case the integration can be carried out 
exactly and results in
\begin{equation}
\begin{array}{c}
\rho^{osc}= \sum_{\bf m}^{'} \;   \frac{2 (-1)^{m_1+m_2}
e^{\frac{i \pi E}{\hbar \Omega_1 \Omega_2}
(m_2 \Omega_1+m_1 \Omega_2)}}
{\pi \hbar (m_2 \Omega_1-m_1 \Omega_2)} \\ \\
\times \sin{[\frac{\pi E}{\hbar \omega_1 \omega_2}
(m_2 \Omega_1-m_1 \Omega_2)]} 
\end{array} \label{osc}
\end{equation}
Substituting the above expression  for $\rho^{osc}$ in the formulas for 
$\Delta F^1$ and $\Delta F^2$ we get
\begin{equation}
\begin{array}{c}
\Delta F^1= \sum_{\bf m}^{''} \; \frac{(-1)^{m_1+m_2}}
{\hbar^2 \omega_1 \omega_2(\gamma_2-\gamma_1)} 
\frac{4 \pi {\mu^0}^2}{\beta} \\ \\
\times \left[ \frac{\sin{(\gamma_2 )} (1-\delta_{m_2 0})}{\gamma_2 
\sinh{(\frac{\pi \gamma_2}{\mu^0 \beta})}}
- \frac{\sin{(\gamma_1)}(1-\delta_{m_1 0})}{\gamma_1 
\sinh{(\frac{\pi \gamma_1}{\mu^0 \beta})}}
\right]
\end{array} \label{f1comp}
\end{equation}
and
\begin{equation}
\begin{array}{c}
\Delta F^2 = \frac{1}{2 \rho^0} \left\{
 \sum_{\bf m}^{''} \frac{(-1)^{m_1+m_2}}
{\hbar^2 \omega_1 \omega_2(\gamma_2-\gamma_1)} \frac{4 \pi \mu^0}{\beta} 
\right. \\ \\ \times \left.
\left[ \frac{\cos{(\gamma_2)} (1-\delta_{m_2 0})}
{\sinh{(\frac{\pi \gamma_2}{\mu^0 \beta})}}
- \frac{\cos{(\gamma_1)} (1-\delta_{m_1 0})}
{\sinh{(\frac{\pi \gamma_1}{\mu^0 \beta})}}
\right] \right\}^2 \; .  
\end{array}\label{f2comp}
\end{equation}
\noindent The double prime in the summations means that only
the integers $(m_1,m_2)$ in the upper half plane, minus the negative 
$m_1$ axis, are included. We have also defined 
$\gamma_i=2 \pi \mu^0 m_i/(\hbar \Omega_i)$.

\begin{figure}
\setlength{\unitlength}{1mm}
\begin{picture}(150,80)(0,0)
\put(3,10){\epsfxsize=9cm\epsfbox{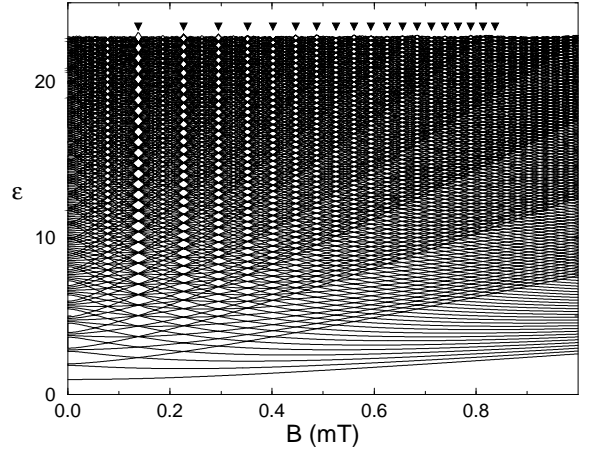}}
\end{picture}
\caption{ The first 290 single particle energy levels normalized to
$\epsilon=E/(\hbar \omega_1)$ as a function of the magnetic field $B$. 
The triangles on top indicate the resonances with $m=1$ and 
$n=2,...,20$.}
\label{fig1}
\end{figure}

Both expressions for $\Delta F^1$ and $\Delta F^2$ can be simplified
if one notes that, as the magnetic field is varied, the ratio 
$\Omega_1/\Omega_2$ passes densely through rational numbers, where all the 
classical orbits of the system are periodic. To those field values there
also corresponds a large degeneracy of the energy levels. We
therefore restrict our attention initially to the neighborhood of these 
values of $B$ only. Let $B=B_{nm}$ be such that $\Omega_1=n \Omega_0$ and
$\Omega_2=m \Omega_0$, i.e,  $\Omega_1/\Omega_2=n/m$ and 
$\Omega_0=\sqrt{\frac{\omega_1 \omega_2}{nm}}$ is the 
frequency of the classical periodic orbits. 
At those points the denominator in (\ref{f1comp}) and (\ref{f2comp}) 
vanishes for all $(m_1,m_2)$ of the form $(pn,pm)$ for all $p > 0$ and 
we see that the main contributions to $\Delta F^1$ and $\Delta F^2$ come 
from these {\em resonant} terms. Defining the function
\begin{equation}
S(B) = \frac{\sin{[C(B)]}}{C(B)}
\end{equation}
where $C(B)=\frac{1}{2}(\gamma_2-\gamma_1)=
\frac{\pi \mu^0}{\hbar \omega_1 \omega_2}(n \Omega_2 - m \Omega_1)$
vanishes at $B_{nm}$, we get, after rearranging the trigonometric functions 
and considering only the term  $p=1$, the following approximated expressions:
\begin{equation}
\Delta F^1 = \frac{(-1)^{n+m} \mu^0  R(\beta)}{\pi^2 n m } \;
\cos{(\frac{2 \pi \mu^0}{\hbar \Omega_0})} \; S(B)
\label{f1res}
\end{equation}
and
\begin{equation}
\Delta F^2 = \frac{\mu^0 R^2(\beta)}{\pi^2 n m } \;
\sin^2{(\frac{2 \pi \mu^0}{\hbar \Omega_0})} \; S^2(B)
\label{f2res}
\end{equation}
\noindent where $R(\beta)=(2 \pi^2)/(\hbar \beta \Omega_0) \;
\sinh^{-1}{[(2 \pi^2)/(\hbar \beta \Omega_0)]}$ is
a temperature dependent factor that diminishes exponentially the 
susceptibility for large $T$'s {\em and} large $\Omega_0$'s. 
The dependence of the free energy on the magnetic field has been
reduced to $S(B)$ and the magnetic susceptibility
can be readily computed. We recall that the Landau susceptibility
for the oscillator is given by \cite{prado} 
$\chi_L = -(e/m^*)^2 \hbar^2/(6 \mu^0)$. 

\begin{figure}
\begin{center}
\vspace{-0.8cm}
\setlength{\unitlength}{1mm}
\begin{picture}(150,80)(0,0)
\put(3,10){\epsfxsize=8.5cm\epsfbox{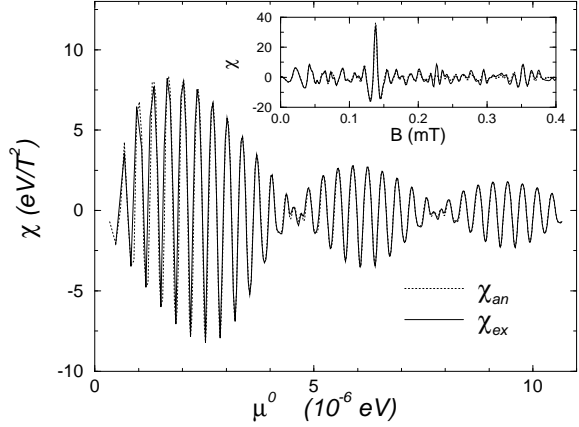}}
\end{picture}
\vspace{-0.5cm}
\caption{Magnetic susceptibility $\chi$ as a function of 
$\mu^0$ at $B=0$ and $T=0.3$ mK. The inset shows $\chi$ as a function of 
B for a single system, without any average, for the same temperature and 
$N=500$. In both cases the full line represents the numerical exact result 
and the dotted line shows the analytical calculation  
$(\chi^L+\chi^1+\chi^2)$.}
\label{fig2}
\end{center}
\end{figure}

The expressions just derived describe the behavior of a single system
in the thermodynamic limit. In the experiment with square billiards of 
Levy et al \cite{levy}, however, only the average properties of an ensemble of 
systems were measured. The individual members of the ensemble, although very
similar, present small differences among themselves. Besides, the
number of particles confined in each of them might vary slightly. To account 
for these fluctuations further averages have to be performed 
\cite{ullmo,report}. As in the case of billiards \cite{ullmo} the oscillatory
contribution of $\Delta F^1$ to $\chi$ vanishes under an average over the 
Fermi energy, or number of particles, for dispersions $\delta \mu$ of the 
order of $\hbar \Omega_0$ . The contribution of $\Delta F^2$ 
remains for the parabolic potential as it does in the case of billiards. 
The main difference here is that, due to the fact that the density of states 
increases linearly with the energy, the contribution of $\Delta F^2$ is of the 
same order in $\mu^0$ than that of $\Delta F^1$. Also, in terms
of particle number, the relative dispersion $\delta N/N$ necessary
to kill $\Delta F^1$ falls as $1/\sqrt{N}$. Therefore, for large $N$'s,
even very small dispersions will effectively wash out $\Delta F^1$. The 
resulting susceptibility, after performing the average, is
\begin{equation}
\langle \chi \rangle_N = -\frac{\mu^0}{2 \pi^2 N n m } \;
R^2(\beta) \; \frac{\partial^2 S^2(B)}{\partial B^2} \; \; .
\label{chin}
\end{equation}
Therefore, since $\partial^2 S^2(B)/\partial B^2$ is proportional to
$N$ and has a negative peak at $B_{nm}$, $\chi$ exhibits a positive peak at 
each resonance whose strength goes as $\mu^0/nm$.

\begin{figure}
\setlength{\unitlength}{1mm}
\begin{picture}(150,80)(0,0)
\put(14,10){\epsfxsize=6.0cm\epsfbox{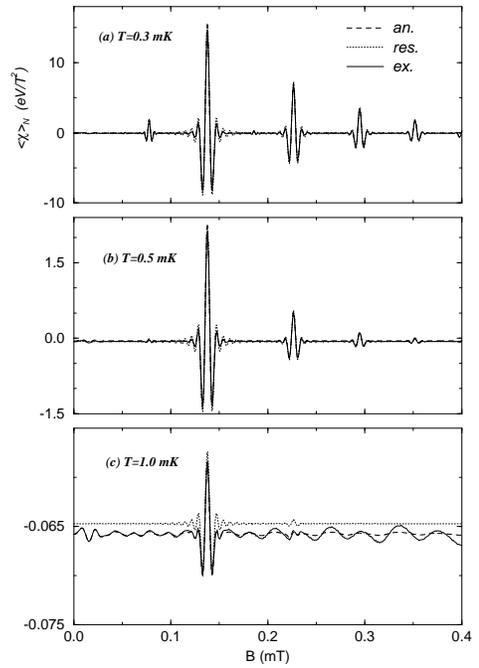}}
\end{picture}
\caption{Susceptibility averaged over $N$ as a function of
the magnetic field for three different temperatures. In all
cases $N=500$, $\delta N/N = 0.2$, the full line represents the exact result, 
the dotted line shows the resonant approximation and the dashed line shows
the full analytic formula, Eq.(\ref{f1comp},\ref{f2comp}) }
\label{fig3}
\end{figure}

In what follows we present numerical calculations performed with 
$\omega_1=5.4 \times 10^{8} s^{-1}$, $\omega_2=0.9 \omega_1$ and
$m^* = 0.067 m_e$, which is the electron effective mass for a GaAs quantum
well. Fig.\ref{fig1} shows the first 290 energy
levels as a function of the magnetic field $B$. The arrows on top indicate
the position of the most relevant resonances.  Fig.\ref{fig2} shows  $\chi$ 
as a function of $\mu^0$ at $B=0$ for a single system, without the average. 
The inset shows $\chi$ as a function of $B$ for $500$ particles, corresponding
to approximately $\mu^0=7.5 \times 10^{-6}$ eV. In both cases $T=0.3$mK, the 
full line represents the numerical exact result and the dotted line shows 
the result derived from Eqs.\ref{f1comp} and \ref{f2comp}. The agreement 
between exact and analytic results is very good. The approximate
formulas, Eqs.\ref{f1res} and \ref{f2res} also give very accurate results
for $\chi$ close to the resonances. Fig.\ref{fig3} shows  
$\langle \chi \rangle_N$, the susceptibility averaged over particle number, 
as a function of $B$ for $N=500$ and dispersion $\delta N/N = 0.2 $ computed 
directly from the energy levels (full line), 
from the resonant formula (\ref{chin}) (dotted line) and from the
full analytic expression Eqs.(\ref{f1comp},\ref{f2comp}) (dashed line) for 
three different values of the temperature. The last two curves also include 
the contribution of $\chi_L$. Once again we found 
an excellent agreement between the exact and approximate calculations (notice
that the exact and analytical curves involve the separate calculation of
$\chi$ for various $N$'s before the average is performed).

\begin{figure}
\setlength{\unitlength}{1mm}
\begin{picture}(150,80)(0,0)
\put(17,10){\epsfxsize=6.0cm\epsfbox{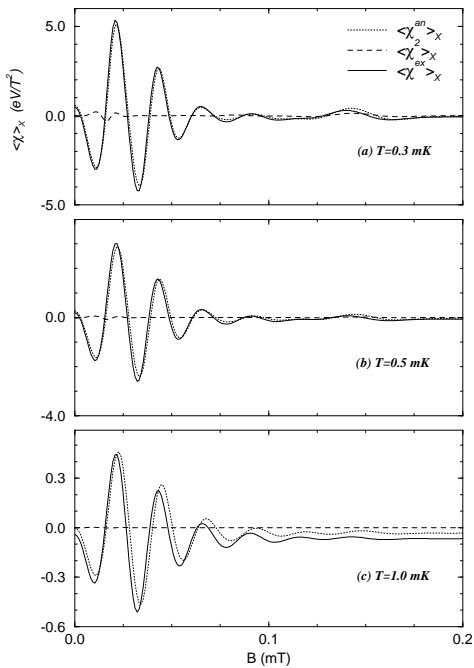}}
\end{picture}
\caption{{\em Size} averaged susceptibility (see text) as a function of
the magnetic field for three different temperatures. In all
cases $\delta x = 0.1$, the full line represents the exact 
result and the dotted and dashed lines show the contributions of the
full analytic calculation and that of $\chi^2$ alone respectively.}
\label{fig4}
\end{figure}

We finally consider averages over different confining potentials. In order
not to introduce too many parameters we define each member of the ensemble
to have $\omega_1=x \omega_{10}$ and 
$\omega_2=x \omega_{20}$, with a gaussian distribution of $x$ around 
$\bar{x}=1$. This can be considered as a {\em size} average, since $x$
changes the available area in coordinate space without changing the
shape of the potential. In what follows we use
an extra index $0$ to indicate quantities computed with $x=1$. Keeping the 
number $N$ of particles fixed we see that
$\mu^0=\hbar \sqrt{N x^2 \omega_{10} \omega_{20}} = x \mu^0_0$ and
$\Omega_0(B)=x \Omega_{00}$. Therefore, the oscillations in (\ref{f1res})
and (\ref{f2res}), which depend on the ratio $\mu^0/\Omega_0$,  are not 
affected by the average. However, since $\Omega_i(B)=x \Omega_{i0}(B/x)$,
$S(B)=S_0(B/x)$ and averaging over $x$ is equivalent to average over $B/x$.
Writing $x=1+\delta x$, $B/x \sim B- B \delta x$ and we see that the average
is not effective for small values of the magnetic field. 
Therefore, the resonant peaks at large $B$'s tend to be smoothed
out, enhancing the susceptibility at the non-resonant 
region, close to $B=0$ for the current value of the parameters. The 
non-resonant regions are described approximately by 
Eq.(\ref{f1comp}) with $(m_1,m_2)=(1,1)$ and
show an oscillatory behavior with frequencies $\gamma_1$ and $\gamma_2$.
Expanding $\gamma_i(B/x) \sim \gamma_i(B) -B \delta x \gamma_i^{'}$
and imposing $|\gamma_i(B/x) - \gamma_i(B)| =2 \pi$ we find that the
oscillations in $\chi$ die out for
$e B/m^* \sim \sqrt{\hbar \bar{\Omega} (\omega_1^2-\omega_2^2)/
(\mu^0_0 |\delta x|)}$ where $\bar{\Omega}$ is the smallest between 
$\Omega_1$ and $\Omega_2$. This is confirmed by the numerical data displayed 
in Fig.\ref{fig4} for  $\delta x = 0.1$ and different temperatures. Notice 
that the contribution of $\Delta F^2$ can be neglected because it falls off 
much faster with the temperature than $\Delta F^1$. 

In conclusion, the magnetic response of an ensemble of two-dimensional 
electron gases confined by parabolic potentials is qualitatively different from
that of an ensemble of billiards. The main features of the problem can
be understood with the help of a {\em resonant} approximation for
the density of states. For an ensemble of identical quantum wells, each holding
slightly different number of electrons, the magnetic response is enhanced only
at the resonances. When a dispersion in the size of the oscillators is 
is included, the sharp response at the resonances are smoothed out but the
susceptibility at low fields remains oscillatory as a function of the Fermi
energy, and not necessarily paramagnetic as in the case of billiards. We 
emphasize that the nature of our approximations 
are different from those of ref.\cite{ullmo}, since here it is validy for
all values of $B$, not only in the limit of small fields.  
The large peaks exhibited by the 
susceptibility at the resonances, is very peculiar of the oscillator
but the dependency of $\rho^0$ with the energy, on the other hand, is a
generic property on smooth confinements and plays an important role
in balancing the relative contributions of $\Delta F^1$ and $\Delta F^2$ at low
temperatures. We notice that the oscillator parameters and magnetic field can
be scaled in order to allow for experimentally accessible values. If
the frequencies are both multiplied by $f$, the Fermi energy, 
the susceptibility and the magnetic field are also multiplied by $f$,
whereas the density of states scales as $1/f$. For $f=10^3$, for 
instance, we would still be considering fields of the order of $0.1$
Tesla.

%%%%%%%%%%%%%%%%%%%%%%%%%%%%%%%%%%%%%%%%%%%%%%%%%%%%%%%%%%%%%%%%%%%%%%%%%%
%%%%%%%%%%%%%%%%%%%%%%%%%%%%%%%%%%%%%%%%%%%%%%%%%%%%%%%%%%%%%%%%%%%%%%%%%%
\acknowledgments

This work was partially supported by the Brazilian agencies Funda\c{c}\~ao
de Amparo \`a Pesquisa do Estado de S\~ao Paulo (FAPESP) and 
Financiadora de Estudos e Projetos (FINEP).

%%%%%%%%%%%%%%%%%%%%%%%%%%%%%%%%%%%%%%%%%%%%%%%%%%%%%%%%%%%%%%%%%%%%%%%%%%
%%%%%%%%%%%%%%%%%%%%%%%%%%%%%%%%%%%%%%%%%%%%%%%%%%%%%%%%%%%%%%%%%%%%%%%%%%

\end{document}